# Why Neutrino Lines are Hypersharp [a]

R. S. Raghavan
Department of Physics and IPNAS, Virginia Tech Blacksburg VA 02460

It was recently pointed out that mono-energetic neutrino lines from the 2-body decay of tritium ($\tau$~18-y) can be emitted, a significant fraction, with natural line width (~$10^{-24}$ eV) for hypersharp resonance transitions $^3$H↔$^3$He. The very long lifetime typical of a neutrino transition, encountered for the first time in resonance, is the key to this surprising effect which is not intuitive from perspectives of line broadening in resonances of short lived ($\tau$~μs) states

Monoenergetic neutrinos ($\tilde{\nu}_e$) from the two-body decay of tritium (T) can induce resonant reactions $^3$H↔$^3$He in a manner similar to γ-rays in the familiar Mőssbauer Effect (ME) [1]. In a recent paper [2] I have argued that this resonance sets a unique framework in which the $\tilde{\nu}_e$ are *hypersharp*, i.e. with the extremely narrow natural width (~$10^{-24}$ eV) of the 18-y T state, thus capable of extremely high resonance cross section approaching $\sigma$ ~$10^{-20}$ cm$^2$.

Comments by Potzel and Wagner[3] and Schiffer[4] seek to contest these conclusions. Their perspective is classical ME experience drawn exclusively from short lived states. They miss the key role of the long nuclear lifetime in the tritium $\tilde{\nu}_e$ resonance in which energy fluctuations *modulate* rather than *detune* the resonance. The averaging effect of modulation is ubiquitous in physics (indeed it forms the physics basis of the ME itself [5]). The long nuclear lifetime, encountered for the first time in resonance transitions, highlights a unique effect that relegates all line broadening to well-resolved *sidebands* instead of detuning the central line. As a result, a significant fraction of the "carrier" $\tilde{\nu}_e$ has *no choice* but to be emitted with the natural line width. Ignoring this feature leads to the misleading conclusions of (3,4).

***Periodic perturbations:*** We begin with a major common ground with (3) which agrees that *periodic* fluctuations *do* indeed result in motional averaging leading to the natural line width for tritium (T) $\tilde{\nu}_e$. This concession is crucial in recognizing the difference due to long and short lifetimes in the response to energy fluctuations. For long lifetimes the $\tilde{\nu}_e$ line is modulated (not just detuned) by fluctuations with two parameters[6] $\eta = (\Omega_o/\Omega)$, the ratio of the energy spread to the fluctuation frequency that affects the central line *intensity* and $\xi = (\Omega/\Gamma)$ the separation from the central line in linewidths $\Gamma$ of the sidebands created by the modulation.[2] Since in the long lived T case ($\Gamma$ extremely small) $\xi$ is always >> 1 for conceivable fluctuations in solids the sidebands are *always* resolved from the central line by many orders of magnitudes of linewidths. Thus a significant fraction of the emitted $\tilde{\nu}_e$ line is *necessarily of natural width*. An acceptable fraction is lost to the sidebands given by the Bessel coefficient $[J_o(\eta)]^4$ for the complete process $^3$H↔$^3$He. As the lifetime becomes shorter, ~μs in usual ME cases, $\Gamma$ is many orders of magnitude larger, $\xi$ decreases to < 1, the sideband resolution is lost and the central line appears broad for the same fluctuation parameters

Periodic energy fluctuations are pervasive in solids because of lattice vibrations. They not only average the Doppler effect due to atomic motion and create recoilless emission[5], but extend the effect to fluctuations of energies of all coordinate (r) dependent interactions. The principle applies equally to *mechanical* vibrations and ultimately, perhaps to gravitational waves[7], the universal periodic perturbation.

***Stochastic perturbations:*** Comment (3) now asserts that line narrowing as above does not result from *stochastic* fluctuation. This directly contradicts well known works such as that of Dicke[8] who showed that even the Doppler profile of spectral lines due to stochastic collisions in a *gas* manifests a sharp unshifted line, indeed anticipating the ME by many years. The fluctuation spectrum due to random relaxation process of (3) was shown by Kubo[9] to be a Gaussian. It is nearly so also for a relaxing two level system cited by (3). The components of the Gaussian modulate the $\tilde{\nu}_e$ energy in the same way as those of a complex spectrum of periodic lattice vibrations in a solid. As a result, the broad modulating Gaussian appears in the *side bands,* not the central line. The Gaussian sideband with the mean energy and width (many orders of magnitude wider than the natural width) is still well resolved from the sharp central line because $\xi >>1$ due to the long T lifetime. The only effect is some loss of central line intensity. Using the same relaxation parameters as (3), the hypersharp $\tilde{\nu}_e$ intensity is $[J_o(\eta)]^4$ for $\eta = [(6.6x\ 10^{-11}/5x10^{-11}) \sim 1.23] \sim 2x10^{-1}$, not $\sim 10^{-11}$ as claimed by (3).

***Inhomogeneous perturbations*** (3,4)*:* Random inequivalent sites create a distribution of perturbing fields, an important consideration. An example is electric interaction due to random lattice defects which is absent in our case since the nuclear quadrupole moments of T and $^3He$ are zero. Inhomogeneous dipolar fields ($\propto \mu_1 \mu_2/r^3$) of neighboring atoms are however, present. The dipolar field is not static but fluctuates typically by 0.02 gauss[2] (by displacements in lattice vibrations). The energy spread in the fluctuation is $\Delta E = \Omega_o \sim 2x10^{-13}$ eV which *modulates* the $\tilde{\nu}_e$ energy (not detunes as feared in (3)) with the frequency of the zero point motion (ZPE) of ~0.1 eV. The intensity of the hypersharp $\tilde{\nu}_e$ is preserved since $[J_o(\eta)]^4 \sim 1$ with $\eta \sim 10^{-13}/0.1$. Balko et al[10] showed that in *general* inhomogeneous fields are averaged out by fast relaxation (a revisit of this work with explicit inclusion of the nuclear lifetime as done in (6) is desirable).

***Final State Energy Differences:*** The T experimental design is based on a unique metal tritide host in which, the T and He sit in well- defined isosymmetric tetragonal interstitial traps.[11] The atomic dynamics of T and He are controlled by *discrete* vibrational excitations in the local potential wells. The recoilless fraction is determined only by these local excitations (known from neutron inelastic scattering[12]), not the Debye spectrum of the bulk lattice.

A new aspect specific to hypersharp emission in T ↔He transition (in addition to motional averaging) is the role of final state energy compensation. The $\tilde{\nu}_e$ energy is changed in emission and absorption by final state energies of atomic shells and trap binding and dynamics (such as ZPE). Each of these is different for T and He. In each emission and absorption *both* T and He participate. Thus the net final state energy change Δ in the decay T➔He changes sign in the absorption He➔T and cancelled in the complete process $^3H \leftrightarrow ^3He$. Further, all r-dependent energies such as the local well depths, binding energies and ZPE's are averaged to *unique hypersharp* central values. Self compensation is thus precise and the hypersharp resonance condition is strictly maintained.

***Temperature dependence*** *(4):* The discrete excitation energies in the potential wells are known from neutron inelastic scattering[12]. The lowest vibrational level is at 72 meV, equivalent to ~850K, below which the atomic motions cannot excite a trap level. Thus at normal temperatures the hypersharp fraction is largely independent of temperature. There is thus little temperature dependent Josephson shift[13] either, only a shift in the constant ZPE. As discussed above, the ZPE shift changes sign in the complete process and cancelled. These aspects differ sharply from the temperature dependence of the classical ME due to excitation of low frequencies in the bulk lattice Debye spectrum. The experience in maintaining ultra small temperature gradients in ton scale ultracold bar detectors of gravitational waves should be very helpful in eliminating residual temperature dependent shifts in gram scale masses.

***Lattice Excitations due to Static Dilation ?***: The formation of the trap is a local dilation extending at the most to the next two nearest neighbors. The dilation creates at the same time, local vibrational levels. The $\tilde{\nu}_e$ energy is changed by the trap energy (as discussed above) and possible excitation of states in the trap, also taken into account in the estimation of the resonance rates. For the dilation to also additionally excite states in the bulk lattice outside the trap (as suggested by (3)), the shock wave due to the dilation must sample the bulk. This can occur only at the speed of sound, thus long after $\tilde{\nu}_e$ emission. Note that this effect does *not affect the natural line width*-- only the recoilless probability. Thus even with the highly unlikely effect, the resonance cross section is still huge, $\sigma \sim 10^{-26}$ $cm^2$, adequate for most of the physics applications.

***Mechanical Vibrations:*** Because of the extreme sharpness at issue it is legitimate to raise the question[4] of the effect of mechanical stability and vibrations on the resonance via the Doppler effect . The latter can be estimated as: $\delta E/E = v/c = 1/c\ (\delta l/l)\ f\ L = 3x10^{-11} x\ 10x10^{-17} \sim 3x10^{-27}$, where L is the baseline (~10 cm) and $(\delta l/l)$ the fractional length change at the frequency f, taken to be : $(\delta l/l)f \sim 10^{-17}$ Hz from the data of the LIGO gravitational wave detector with a baseline of several km (thus much more demanding than our bench scale problem). Thus the vibration effect is $\delta E/E \sim 3x10^{-27}$ or ~60 linewidths. This appears not as resonance detuning but a periodic *modulation* with $\eta = (\Omega_o/\Omega) = \delta E/(\hbar f.2\pi)$ which ranges from $10^{-5}$ to $10^{-11}$ for frequencies f from $10^{-3}$ to $10^{3}$ Hz. The hypersharp fraction $[J_o(\eta)]^4$ is thus close to unity in the entire range even for mechanical instabilities $(\delta l/l)$ some 5 orders of magnitude larger. The sidebands with the small residual intensity occur at $\delta E/\Gamma \sim 60$ linewdths away from the central line. Indeed, the result suggests external modulation for averaging out fluctuations in practical geometries similar to mechanical vibrations.

***Time- Energy Uncertainty*:** If the dwell time of instability of a decaying state is uncontrolled, the energy spread of the $\tilde{\nu}_e$ is the natural width. If the dwell time of the state before decay is known, the "age" of the source << the lifetime, forces the width to be broader than the natural width. Thus the signal at tritium resonance spontaneously grows with time as the source ages. The unique "age" effect is discussed recently in detail [14] and noticed also elsewhere [15] (refuting recent remarks on the question by Lipkin[16] ).

---

[a]Phys. Rev. Letters in Press (Reply to (ref. 3,4) in a slightly abridged form). The material was presented at NUTECH 09 at ICTP (July 2009).